\documentclass[12pt]{article}

\newcommand{\be}{\begin{equation}}
\newcommand{\ee}{\end{equation}}
\newcommand{\bel}[1]{\begin{equation}\label{#1}}
\newcommand{\bea}{\begin{eqnarray}}
\newcommand{\eea}{\end{eqnarray}}
\newcommand{\ba}{\begin{array}}
\newcommand{\ea}{\end{array}}

\newcommand{\ket}[1]{\mbox{$| \, {#1}\, \rangle$}}
\newcommand{\exval}[1]{\mbox{$\langle \, {#1}\, \rangle$}}

\def\bbbc{{\mathchoice {\setbox0=\hbox{$\displaystyle\rm C$}\hbox{\hbox
to0pt{\kern0.4\wd0\vrule height0.9\ht0\hss}\box0}}
{\setbox0=\hbox{$\textstyle\rm C$}\hbox{\hbox
to0pt{\kern0.4\wd0\vrule height0.9\ht0\hss}\box0}}
{\setbox0=\hbox{$\scriptstyle\rm C$}\hbox{\hbox
to0pt{\kern0.4\wd0\vrule height0.9\ht0\hss}\box0}}
{\setbox0=\hbox{$\scriptscriptstyle\rm C$}\hbox{\hbox
to0pt{\kern0.4\wd0\vrule height0.9\ht0\hss}\box0}}}}
\def\bbbq{{\mathchoice {\setbox0=\hbox{$\displaystyle\rm
Q$}\hbox{\raise
0.15\ht0\hbox to0pt{\kern0.4\wd0\vrule height0.8\ht0\hss}\box0}}
{\setbox0=\hbox{$\textstyle\rm Q$}\hbox{\raise
0.15\ht0\hbox to0pt{\kern0.4\wd0\vrule height0.8\ht0\hss}\box0}}
{\setbox0=\hbox{$\scriptstyle\rm Q$}\hbox{\raise
0.15\ht0\hbox to0pt{\kern0.4\wd0\vrule height0.7\ht0\hss}\box0}}
{\setbox0=\hbox{$\scriptscriptstyle\rm Q$}\hbox{\raise
0.15\ht0\hbox to0pt{\kern0.4\wd0\vrule height0.7\ht0\hss}\box0}}}}
\def\bbbt{{\mathchoice {\setbox0=\hbox{$\displaystyle\rm
T$}\hbox{\hbox to0pt{\kern0.3\wd0\vrule height0.9\ht0\hss}\box0}}
{\setbox0=\hbox{$\textstyle\rm T$}\hbox{\hbox
to0pt{\kern0.3\wd0\vrule height0.9\ht0\hss}\box0}}
{\setbox0=\hbox{$\scriptstyle\rm T$}\hbox{\hbox
to0pt{\kern0.3\wd0\vrule height0.9\ht0\hss}\box0}}
{\setbox0=\hbox{$\scriptscriptstyle\rm T$}\hbox{\hbox
to0pt{\kern0.3\wd0\vrule height0.9\ht0\hss}\box0}}}}
\def\bbbs{{\mathchoice
{\setbox0=\hbox{$\displaystyle     \rm S$}\hbox{\raise0.5\ht0\hbox
to0pt{\kern0.35\wd0\vrule height0.45\ht0\hss}\hbox
to0pt{\kern0.55\wd0\vrule height0.5\ht0\hss}\box0}}
{\setbox0=\hbox{$\textstyle        \rm S$}\hbox{\raise0.5\ht0\hbox
to0pt{\kern0.35\wd0\vrule height0.45\ht0\hss}\hbox
to0pt{\kern0.55\wd0\vrule height0.5\ht0\hss}\box0}}
{\setbox0=\hbox{$\scriptstyle      \rm S$}\hbox{\raise0.5\ht0\hbox
to0pt{\kern0.35\wd0\vrule height0.45\ht0\hss}\raise0.05\ht0\hbox
to0pt{\kern0.5\wd0\vrule height0.45\ht0\hss}\box0}}
{\setbox0=\hbox{$\scriptscriptstyle\rm S$}\hbox{\raise0.5\ht0\hbox
to0pt{\kern0.4\wd0\vrule height0.45\ht0\hss}\raise0.05\ht0\hbox
to0pt{\kern0.55\wd0\vrule height0.45\ht0\hss}\box0}}}}
\def\bbbz{{\mathchoice {\hbox{$\sf\textstyle Z\kern-0.4em Z$}}
{\hbox{$\sf\textstyle Z\kern-0.4em Z$}}
{\hbox{$\sf\scriptstyle Z\kern-0.3em Z$}}
{\hbox{$\sf\scriptscriptstyle Z\kern-0.2em Z$}}}}

\def\gsim {\mbox{\hbox{ \lower-.6ex\hbox{$>$}
\kern-1.12em \lower.5ex\hbox{$\sim$}\kern+.35em}}}
\def\lsim {\mbox{\hbox{ \lower-.6ex\hbox{$<$}
\kern-1.12em \lower.5ex\hbox{$\sim$}\kern+.35em}}}

\begin{document}

\begin{center}          
{\large                       
{\bf Microscopic structure of travelling wave solutions in a class of
stochastic interacting particle systems}
}\\[3cm]
{\large {\sc
K. Krebs$^\ast$, F.H. Jafarpour$^\dagger$, and G.M. Sch\"utz$^\ddagger$
}}\\[8mm]
{\em $^\ast$ Physikalisches Institut, Universit\"at Bonn, 53111 Bonn,
Germany\\
$^\dagger$ Department of Physics, Bu-Ali-Sina University, Hamadan, Iran\\
and Institutes for studies in theoretical Physics and Mathematics (IPM)
School of Physics, P.O. Box 19395-5531, Tehran, Iran\\
$^\ddagger$ Institut f\"ur Festk\"orperforschung, Forschungszentrum J\"ulich,\\
52425 J\"ulich, Germany
}
\vspace{1cm}\\                %
\begin{minipage}{13cm}{
\baselineskip 0.3in
We obtain exact travelling wave solutions for three families of stochastic
one-dimensional nonequilibrium lattice models with open boundaries.
These solutions describe the diffusive motion and microscopic structure
of (i) of shocks in the partially asymmetric exclusion process with open
boundaries, (ii) of a lattice Fisher wave in a reaction-diffusion system, and
(iii) of a domain wall in non-equilibrium Glauber-Kawasaki dynamics with
magnetization current. For each of these systems we define a microscopic shock
position and calculate the exact hopping rates of
the travelling wave in terms of the transition rates of the microscopic model.
In the steady state a reversal of the bias of the travelling wave marks a
first-order  non-equilibrium phase transition, analogous to the Zel'dovich
theory of kinetics of first-order transitions. The stationary distributions
of the exclusion process with $n$ shocks
can be described in terms of $n$-dimensional representations of matrix
product states.\\[4mm]
PACS numbers: 05.70.Ln, 82.20.Mj, 82.40.Fp, 02.50.Ga
}
\end{minipage} \end{center} 

\newpage
\baselineskip 0.3in
\section{Introduction}\label{intro}
Systems of diffusing and reacting particles are usually described on the
macroscopic level by hydrodynamic equations for coarse-grained quantities like
the particle density which represent the order parameter specifying the
macroscopic state of the system \cite{Fife79}. 
Paradigmatic examples for these equations are
the Burgers equation for driven diffusive systems with particle conservation
\cite{Burg74} or the Fisher
equation for reactive  systems without conservation law \cite{Fish37,Lebo88}.
These
equations are in general non-linear and exhibit shocks in some cases. This
means that the solution of the macroscopic equations may develop a
discontinuity even if the initial particle density is smooth.
In order to understand the emergence of such
behaviour from the microscopic laws that govern the stochastic
motion and interaction of
particles it is necessary to {\it derive} the macroscopic equations
from the microscopic dynamics rather than postulating them on phenomenological
grounds. To solve this problem it is evident that detailed insight in the 
microscopic
structure of non-equilibrium systems exhibiting macroscopic discontinuities
must be obtained.

A considerable body of results of this nature has been obtained for specific
one-dimensional lattice models defined on the integer lattice $\bbbz$
\cite{Spoh91,Reza91,Kipn99}, the best-studied example being the asymmetric
simple exclusion process (ASEP) \cite{Ligg99,Schu00}. In this
basic model for a
driven diffusive system each site $k$ is either
empty  ($n_k=0$) or occupied by at most one particle ($n_k=1$). A particle on
site $k$ hops randomly to the site $k+1$ with rate $D_r$  and to the site
$k-1$ with rate $D_l$, but only if the target site is empty.
Otherwise the attempted move is rejected. The jumps occur independently in
continuous time with an exponential waiting time distribution.
 In the hydrodynamic limit the
system is described by the Burgers equation which exhibits
shocks. Such a shock discontinuity may be viewed as the interface between
stationary domains of different densities.
Relaxing the requirement
of particle number conservation leads to class of systems which are
generically called reaction-diffusion processes, but which by a variety
of mappings also serve as toy models for non-conservative spin-flip dynamics
(in the context of magnetic systems), epidemic spreading, growth
processes and transport phenomena in biological and ecological
systems and elsewhere \cite{Schu00,Schu03}.

Most of the results for the dynamical behaviour have been obtained for
infinite particles systems. In many of the physical applications, however,
one has to study finite systems with
open boundaries where particles are injected and extracted.
This is crucial to take into account as -- in the absence of
equilibrium conditions --
the boundary conditions determine the bulk behavior of
driven systems, even to the extent that
boundary induced phase transitions between bulk states of different densities
occur \cite{Krug91,Schu93,Derr93}.
Qualitatively, the strong effect of boundary conditions on the bulk can be
attributed to the presence of steady-state currents which carry boundary
effects
into the bulk of the system. Quantitatively, exact results for the
steady state of the ASEP have helped to show that
part of the nonequilibrium phase diagram of driven diffusive systems with
open boundaries, viz. phase transitions of first order, can be understood
from the diffusive motion of shocks \cite{Kolo98,Popk99}, in analogy
to the Zel'dovich theory of equilibrium kinetics of first-order transitions.
As in equilibrium, the nonequilibrium theory of boundary-induced
phase transitions requires the existence of shocks which are microscopically
sharp.

In a series of recent papers \cite{Parm03} - \cite{Rako03} these
considerations, originally formulated for conservative dynamics, have
been extended to non-conservative reaction-diffusion systems.
Moreover, there are exact results about shocks in
reaction-diffusion systems with branching and coalescence
\cite{Doer88}-\cite{RioDoebAv1995} (here shocks are known as
Fisher waves on the macroscopic scale) and in spin-flip systems where shocks
correspond to domain walls \cite{Glau63}.
However, no exact
results have been reported so far for non-stationary travelling waves
in open systems.
 Here we wish (i) to establish a
complete picture about exact travelling wave solutions for the specific family
of systems
to which these processes belong (viz. single-species exclusion processes
with two-body nearest neighbour interaction and no internal degrees of freedom)
and (ii) to study the dynamics and
microscopic structure of these travelling shocks in
systems with {\it open boundaries}.
Since many of the powerful techniques used for treating the ASEP
do not apply to nonconservative systems we propose a general approach
that can be applied to any lattice model: We take as initial distribution
a shock distribution with given microscopic properties and determine
the class of models for which the shock distribution evolves into a
linear combination of similar distributions with different
shock positions. In this paper we identify three families of processes with
this property.

The paper is organized as follows: In the following section we define
the class of models that we consider and we also define shock measures
for these systems. In Sec.~3 we determine the families of models
with travelling wave solutions on the finite lattice. This is followed by
some new results for the ASEP with open boundaries in Sec.~4.
In Sec.~5 we summarize our results and draw some conclusions.

\section{Reaction-diffusion systems and shock measures}
\label{Sec2}

\subsection{Stochastic single-species models}

We consider Markovian interacting particle systems of a single
species of particles without
internal degrees of freedom which have hard-core
two-body interactions with their nearest neighbour sites.
We describe hard-core interaction due to excluded volume
in terms of an exclusion processes where each lattice site may be occupied
by at most one particle. This class of models
may therefore be described by a set of occupation numbers
$\underline{n}=\{n_1,\dots,n_L\}$ where $n_k=0,1$ is the number of particles
on site $k$ on a lattice of $L$ sites. There is a one-to-one
correspondence to classical spin systems where the
occupation number $n_k=0$ represents spin up while $n_k=1$ represents
spin down.

The stochastic dynamics are
defined in terms of transition rates (transition probabilities per
infinitesimal time unit). The process is fully defined by the 12 rates
$w_{ij}$ for changes of the
configuration of a pair of neighbouring sites $k$ and $k+1$ \cite{Schu95}:
\begin{equation}
\label{2-1}
w_{ij}: \quad
(n_kn_{k+1}) \rightarrow (n_k'n_{k+1}')
\end{equation}
where $i=1,2,3,4$ is the decimal value plus one of the target configuration
$(n_k'n_{k+1}')$ read as a two-digit binary number and $j$ is the respective
value of the initial configuration $(n_kn_{k+1})$, as shown below.
\begin{eqnarray}
\mbox{Diffusion to the left and right } (01 \rightleftharpoons 10)
   \quad & w_{32}, \; w_{23} \nonumber \\
\mbox{Coalescence to the left and right } (11 \to 10,01)
   \quad & w_{34}, \; w_{24} \nonumber \\
\mbox{Branching to the left and right } (10,01 \to 11)
   \quad & w_{43}, \; w_{42} \nonumber \\
\mbox{Death to the left and right } (10,01 \to 00)
   \quad & w_{13}, \; w_{12} \nonumber \\
\mbox{Birth to the left and right } (00 \to 10,01)
   \quad & w_{31}, \; w_{21} \nonumber \\
\mbox{Pair Annihilation and Creation } (11 \rightleftharpoons 00)
   \quad & w_{14}, \; w_{41} \nonumber
\end{eqnarray}
From time to time we also use the more intuitive symbols
$D_r=w_{23},D_l=w_{32}$ for the hopping rates.
Notice that combinations of individual processes may describe other
physically meaningful processes. E.g. coalescence and death with
equal rates is equivalent to single-site radioactive decay
$(1\to 0)$ with that rate. The
inverse of the rate is the mean life time of a particle.
For injection and extraction of particles at the boundaries we introduce the
rates
\begin{eqnarray}
\mbox{Injection and Extraction at the left boundary }(0 \rightleftharpoons 1)
   \quad & \alpha, \; \gamma \nonumber \\
\mbox{Injection and Extraction at the right boundary } (0 \rightleftharpoons 1)
   \quad & \delta, \; \beta   \nonumber
\end{eqnarray}

The time evolution is defined by a continuous-time master equation for the
distribution $P(n_1, \cdots, n_L;t)$ which we write in terms of the
quantum Hamiltonian formalism \cite{Schu00}.
The distribution is mapped to a probability vector $\ket{P(t)}$ which
contains as components the probabilities $P(n_1, \cdots, n_L;t)$.
The time evolution is generated by the stochastic Hamiltonian $H$
whose matrix elements are the transition rates between configurations.
The Markovian time evolution can then conveniently be cast in the
form of an imaginary time Schr\"odinger equation
\bel{2-11}
\frac{d}{dt}\ket{P(t)} -H \ket{P(t)}
\ee
with the formal solution
\bel{2-12}
\ket{P(t)} = \mbox{e}^{-Ht}\ket{P(0)}.
\ee

The quantum Hamiltonian $H$ for the family of processes defined above
has the structure
\bel{2-14}
H = \sum_{k=1}^{L-1} h_k + b_1 + b_L.
\ee
Here
\bel{2-15}
h_k = - \left( \ba{cccc}
 .& w_{12} & w_{13} & w_{14} \\
w_{21} & . & w_{23} & w_{24} \\
w_{31} & w_{32} & . & w_{34} \\
w_{41} & w_{42} & w_{43} & . \ea
\right)_{k,k+1}
\ee
is the local transition matrix acting nontrivially on sites $k,k+1$.
The diagonal elements are the negative sum of the transitions rates
in the respective column, as required by conservation of probability.
The boundary matrices
\bel{2-16}
b_1 = - \left( \ba{rr}
-\alpha & \gamma \\
 \alpha & -\gamma \ea \right)_1, \qquad
b_L = - \left( \ba{rr}
-\delta & \beta \\
 \delta & -\beta \ea \right)_L
\ee
generate the boundary processes.

The invariant measures \ket{P^\ast} of the process, i.e., the stationary
probability distributions, satisfy the eigenvalue equation
\bel{2-13}
H \ket{P^\ast} = 0.
\ee
We stress that the analogy to quantum mechanics is a formal one, for
details see \cite{Schu00}.

The equations
of motion for the expected local particle density
take the form \cite{Schu95}
\bea
\label{5-1}
\frac{d}{d t} \exval{n_x(t)} & = & A_1 + A_2 + B_1  \exval{n_{x-1}(t)} +
B_2  \exval{n_{x+1}(t)} - (C_1+C_2) \exval{n_x(t)}\\
& &  +D_1 \exval{n_{x-1}(t)n_x(t)} + D_2 \exval{n_{x}(t)n_{x+1}(t)} \nonumber
\eea
with
\bel{5-2}
\ba{llllll}
A_1 & = & w_{21} + w_{41}       &
B_1 & = & w_{23} + w_{43} - w_{21} - w_{41} \vspace{2mm} \\
C_1 & = & w_{12} + w_{32} + w_{21} + w_{41} \hspace{4mm}
                 & D_1 & = & C_1 - w_{23} - w_{43} - w_{14} - w_{34}
\vspace{2mm} \\
A_2 & = & w_{31} + w_{41}       &
B_2 & = & w_{32} + w_{42} - w_{31} - w_{41}  \vspace{2mm}\\
C_2 & = & w_{13} + w_{23} + w_{31} + w_{41} \hspace{4mm}
                 & D_2 & = & C_2 - w_{32} - w_{42} - w_{14} - w_{24} .
\ea
\ee
In analyzing these equations the question arises how to treat the
nonlinearity in the lattice equation, i.e., the
two-point correlator $D_1 \exval{n_{x-1}(t)n_x(t)}
+ D_2 \exval{n_{x}(t)n_{x+1}(t)}$. Calculating its time-derivative
leads to a coupling to three-point correlation functions and
eventually to a hierarchy of equations which is just as untractable
than the master equation itself. Only for some families of models the system
of equations decouples and exact results can be obtained
\cite{Doer88,Schu95,Cohe63,Pesc94,Alim01}.
Therefore here we do not follow this traditional
approach
but rather investigate the time evolution of the measure.

\subsection{Product measures and shock measures}

The stationary distribution of the family of processes
defined above depends on all the transition rates and is not known
in general. On some parameter manifolds, however, the stationary
distributions are simple Bernoulli product measures
\bel{2-17}
P^\ast(\underline{n}) =
\prod_{j=1}^{L}
 \left((1-\rho)\delta_{n_j,0} + \rho \delta_{n_j,1} \right)
\end{equation}
where the probability of finding a particle at each site $k$ is $\rho$ and
independent of the occupation of other lattice sites, i.e., where there
are no correlations.
It is easy to see that $P^\ast$ depends only on the total number
$N=\sum_k n_k$ of
particles in the configuration $\underline{n}$, one has
$P^\ast(\underline{n}) = (1-\rho)^{L-N}\rho^N$.

In the quantum Hamiltonian formulation this distribution is represented
by a tensor state
\bel{2-18}
\ket{P^\ast} = \left( \ba{c} 1-\rho \\ \rho \ea \right)^{\otimes L}
\equiv \ket{\rho}
\ee
The family of processes for which this is an invariant measure can be
determined easily from (\ref{2-13}). One first determines the manifold
of bulk rates $w_{ij}$ such that
\bel{2-19}
h_k \ket{\rho} = A (n_{k+1}-n_{k})\ket{\rho}
\ee
with an arbitrary constant $A$. This yields three equations for
the 12 bulk parameters. The solutions define a manifold of processes
with uncorrelated stationary distributions, provided the system has
periodic boundary conditions. In order to satisfy (\ref{2-13}) for systems
with open boundaries
one determines the boundary parameters by the relations
\bel{2-20}
b_1 \ket{\rho} = A n_1 \ket{\rho}, \quad b_L \ket{\rho} = - A n_L \ket{\rho}.
\ee
For each boundary this is one equation for two rates.
Notice that these relations contain the stationary particle density $\rho$ as
free parameter.

\begin{figure}[h]
\begin{center}
\begin{picture}(120,70)
\put(10,10){\line(0,1){50}}
\put(10,10){\line(1,0){110}}
\put(10,20){\line(1,0){60}}
\put(70,20){\line(0,1){20}}
\put(70,40){\line(1,0){50}}
\put(120,10){\line(0,1){50}}
\put(10,0){\footnotesize 1}
\put(66,0){\footnotesize $k$}
\put(116,0){\footnotesize $L$}
\put(125,40){\footnotesize $\rho_2$}
\put(0,20){\footnotesize $\rho_1$}
\end{picture}
\caption[profig]{Density profile of a Bernoulli shock measure}
\label{profig}
\end{center}
\end{figure}
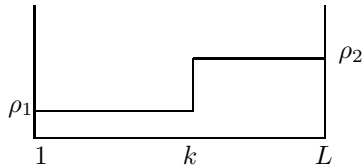

Bernoulli shock measures are product measures with a jump in the density at
some site $m$ (\ref{profig}). They are represented by a tensor state
\bel{2-21}
\ket{k}_{\rho_1,\rho_2} =
\left( \ba{c} 1-\rho_1 \\ \rho_1 \ea \right)^{\otimes k}
\otimes \left( \ba{c} 1-\rho_2 \\ \rho_2 \ea \right)^{\otimes L-k}.
\ee
There are no correlations, but the density in the left domain of sites
$1, \dots, k$ is $\rho_1$ and then jumps to $\rho_2$ in the right domain
$k+1,\dots, L$ of the system. Since there are no correlations one may
regard the lattice unit as the intrinsic shock width. Hence shocks which
are described by a such a distribution are microscopically sharp and have
a very simple internal structure, characterized by the absence of any
correlation between particle positions. There is no process of the form
(\ref{2-14}) for which a shock distribution with a shock at some given
site $k$ is stationary. However, as shown in next section, linear
combinations of shock measures may be stationary distributions.
Notice that a shock at position $m=0$ corresponds to a Bernoulli measure
with density $\rho_2$ while a shock at position $k=L$ corresponds to a
Bernoulli measure with density $\rho_1$.

\section{Shocks as stable collective excitations}

A linear combination of shock measures is a stationary measure if
a given shock measure evolves into
a linear combination of shock measures after time $t$., i.e., where
$\ket{P(t)}_{m,\rho_1,\rho_2} \equiv \exp{(-Ht)} \ket{m}_{\rho_1,\rho_2}$
has the form
\bel{3-1}
\ket{P(t)}_{m,\rho_1,\rho_2} = \sum_{k=0}^{L} p(m,k;t)
\ket{k}_{\rho_1,\rho_2}.
\ee
The physical interpretation of this property is that a shock retains its
internal structure at all times, only the position of the shock is
shifted by a random amount. The probability of moving from the initial
shock position $m$ to site $k$ after time $t$ is the quantity $p(m,k;t)$.
Hence we shall refer to measures of the form (\ref{3-1}) as diffusive
shock measures.
The random walk nature of such a shock gives rise to the interpretation
as a collective single-particle excitation. This can be made more
precise by the implications of (\ref{3-1}). The evolution
into a linear combination of shock measures implies for an infinitesimal
step (which is generated by $H$) the evolution equation
\bel{3-2}
- H \ket{m}_{\rho_1,\rho_2} = d_1 \ket{m-1}_{\rho_1,\rho_2} +
d_2 \ket{m+1}_{\rho_1,\rho_2} - (d_1+d_2) \ket{m}_{\rho_1,\rho_2}
\ee
which is the evolution equation for a biased single-particle random walk
with hopping rate $d_2$ to the right and hopping rate $d_1$ to the left.
Hence we have to determine processes such that (\ref{3-2}) is satisfied.
The family of shock distributions defined by the densities $\rho_1,\rho_2$
forms an invariant sector $\cal U$ under the time evolution of the system.
Notice that (\ref{3-2}) implies the existence of at least two stationary
product solutions in a periodic system. These stationary states have
densities $\rho_1,\rho_2$ respectively.
The boundary rates have to chosen such that at the left boundary $\rho_1$
is stationary, while at the right boundary $\rho_2$ is stationary.

Solving (\ref{3-2}) leads to three classes of reaction-diffusion models
with an invariant sector $\cal U$ which are described below.

\subsection{Asymmetric Simple Exclusion Process (ASEP)}

The simplest process which satisfies (\ref{3-2}) is the ASEP with
hopping to the left and right (without loss
of generality we assume a bias to the right)
and injection and extraction at both boundaries. Hence the nonvanishing rates
\begin{equation}
w_{32},\quad w_{23},\quad \gamma,
\quad \alpha,\quad \beta,\quad \delta.
\end{equation}
These rates together with the densities $\rho_1$ and $\rho_2$ satisfy the
following conditions:
\begin{eqnarray}
w_{23} = \frac{(1-\rho_1)\rho_2}{\rho_1(1-\rho_2)}
w_{32} \label{asepfront1}\\
\alpha(1-\rho_1)  - \gamma \rho_1 = (w_{23}-w_{32})\rho_1(1-\rho_1)
\label{asepfront2} \\
\beta \rho_2 - \delta (1-\rho_2)
= (w_{23}-w_{32}) \rho_2(1 - \rho_2)
\label{asepfront3}
\end{eqnarray}
Both densities have to fullfil the conditions $0 < \rho_1 < \rho_2 < 1$ in
this case. Condition (\ref{asepfront1}) was obtained
for the infinite system in \cite{Beli02}) using symmetry properties
of the quantum Hamiltonian (see next section).
In the bulk of the lattice the shock position moves like a biased
lattice
random walk with hopping rates
\bel{3-3}
d_i = (D_r-D_l) \frac{\rho_i(1-\rho_i)}{\rho_2-\rho_1}
\ee
to left ($i=1$) and right ($i=2$) respectively. This leads to the well-known
exact expressions
for the shock velocity $v_s = d_2-d_1$ and shock diffusion coefficient
$D_s = (d_2+d_1)/2$ \cite{Spoh91,Ferr94} as long as the shock is far
away from the boundaries.

The new result are conditions (\ref{asepfront2}), (\ref{asepfront3})
which imply that at the boundaries the shock is reflected.
According to the properties of a biased random walk on a finite lattice
with reflecting boundaries and bulk rates (\ref{3-3})
its stationary position after equilibration
is geometrically distributed, i.e., the probability of finding the
shock at site $k$ on the lattice is of the form
\bel{3-4}
p^\ast(k) \propto \left(\frac{d_2}{d_1}\right)^k.
\ee

Depending on the bias of  the shock the steady state of the system is in the
low-density subphase $A_I$ (for $d_2>d_1$), in the high-density subphase
$B_I$ (for $d_1>d_2$) or on the first-order coexistence line (for $d_1=d_2$)
\cite{Schu93,Sand94}.
From (\ref{3-4}) we read off the exact inverse localization length
\bel{3-5}
\xi^{-1} = \ln{(d_2)} -  \ln{(d_1)}
\ee
which was conjectured in \cite{Schu93} to describe the localization
of the shock throughout the subphases $A_I$ and $B_I$. In fact, since the
existence of a shock is a generic property of driven diffusive systems, our
results
support the picture developed in \cite{Kolo98} where it is argued that
a localization length of the form (\ref{3-5}) is universal for driven
diffusive systems in the subphases $A_I$ and $B_I$.

\subsection{Branching--Coalescing Random Walk (BCRW)}

In this case we have the following non-vanishing rates:
\begin{equation}
w_{34},\quad w_{24},\quad w_{42},\quad
w_{43},\quad
w_{32},\quad w_{23},\quad
\alpha, \quad \gamma,\quad \beta
\end{equation}
In the periodic system
there are two translation invariant stationary states in
this model: Bernoulli measures with zero density and with a density
$\rho^\ast$ resp. which depends only on the ratio
\be\label{coagfront1}
\frac{w_{24}+w_{34}}{w_{42} + w_{43}}
= \frac{1-\rho^\ast}{\rho^\ast}
\ee
between the branching and coalescence rates. For the existence of diffusive
shock measures in the open system
one of the densities $\rho_i$ has to be zero, the other has to be
$\rho^\ast$. Without loss of generality
we set $\rho_2 = 0$. The non-vanishing rates then have to satisfy
the conditions
\begin{eqnarray}
w_{23} &=&  \frac{1-\rho^\ast}{\rho^\ast}   w_{43}
\label{coagfront2} \\
\gamma &=& \frac{1-\rho^\ast}{\rho^\ast}  \alpha + (1-\rho^\ast)
w_{32}
    - \frac{1-\rho^\ast}{\rho^\ast} w_{43} + \rho^\ast w_{34} .
\label{coagfront3}
\end{eqnarray}
This leaves 7 free parameters.

Instead of the BCRW we could have
chosen a birth-death-diffusion model with rates
obtained from (\ref{coagfront1})-(\ref{coagfront2}) by interchanging
particles and vacancies.
The densities behave under this transformation according to
$\rho_{i} \rightarrow 1-\rho_{i}$ for $i= 1,2$
As microscopic
shock position it is convenient to choose the position of the rightmost
particle.

From (\ref{3-2}) we obtain the shock hopping rates
\bea
d_1 & = & w_{43}/\rho^\ast \\
d_2 & = & w_{43}(1-\rho^\ast)/\rho^\ast.
\eea
Hence the velocity of the shock
\be
v_s = d_2 - d_1 = w_{43}
\ee
is determined by the rate of branching to the right. The diffusion
coefficient reads
\be
D = w_{43}\frac{2-\rho^\ast}{2\rho^\ast} = \frac{w_{43}}{2}\left(1+2
\frac{w_{24}+w_{34}}{w_{42}+w_{43}}\right).
\ee
This quantity remains finite in the limit $\rho^\ast \to 0$.

For a special tuning of the coalescence rates (instantaneous on-site
coalescence which is equivalent to $D_r=w_{24}$, $D_l=w_{34}$) this process
can be solved exactly with the help of the so called inter-particle
distribution
functions (IPDF)  \cite{Cohe63,benA97,Evan84}, or, equivalently, using free
fermion techniques, reviewed in detail in \cite{Schu00}.
By passing to the continuum limit (lattice constant $a\to 0$)
ben-Avraham has shown for the free-fermion choice of coalescence rates with
infinitesimal branching rate (proportional to the lattice constant) that the
model
has shock-like solutions if the initial state has zero density on one side
and stationary density on the other side of the origin \cite{bAv1998a}. He has
has also derived properties of higher order correlations which suggest
the existence of a diffusive shock measure at least for this
special limiting case of rates. Thus one has a picture comparable to the
situation in the asymmetric simple exclusion process, however, with the
difference that in the asymmetric exclusion process the densities on both sides
of the shock are arbitrary while in the coalescence-branching model these
densities are fixed to be zero and $\rho^\ast=\rho_1$ respectively.
Moreover, in the limit of ben-Avraham the stationary density $\rho^\ast$
in the active domain is non-zero, but infinitesimal. Straightforwardly
extending our result
to the infinite system
proves the existence of such a diffusive shock measure and shows that such
a shock solution persists also for finite densities in the active domain.

We remark that for the case of symmetric hopping $D_r=D_l$
interesting macroscopic dynamics arise from (\ref{5-1})
if we consider infinitesimal
rates of branching and
coalescence of the order $a^2$ and also rescale time by $a^2$
(diffusive scale). We set
\be
w_{24} = a^2 \hat{w}_{24},\quad w_{34} = a^2 \hat{w}_{34},\quad
w_{42} = a^2 \hat{w}_{42},\quad w_{43} = a^2 \hat{w}_{43}
\ee
and obtain within mean field theory
\begin{equation}
\label{5-9}
\frac{\partial \rho}{\partial t} =
(D_r+D_l) \frac{\partial^2 \rho}{\partial x^2}
+ \hat{k} \rho (\rho^\ast -\rho )
\end{equation}
with
$\hat{k} = \hat{w}_{24}+ \hat{w}_{34}+\hat{w}_{42}+\hat{w}_{43}$.
This is the usual Fisher equation \cite{Fish37}
which is also known to have travelling wave solutions similar to shocks.

Mean field theory for infinitesimal branching and coalescence rates resp.
is justified since
in this case the dynamics in any finite region are dominated by
the pure exclusion process and hence is expected to be locally stationary
and hence to have no correlations \cite{Popk03}.
Neglecting correlations in the derivation
of the hydrodynamic limit of (\ref{5-1}) for the asymmetric
process and keeping terms up to
second order in the lattice constant one obtains
\begin{equation}
\label{5-10}
\frac{\partial \rho}{\partial t} = \nu \frac{\partial^2 \rho}{\partial x^2}
+ \tilde{v} \frac{\partial \rho}{\partial x} -
\tilde{\theta} \rho \frac{\partial \rho}{\partial x}
+ \tilde{k} \rho (\rho^\ast -\rho )
\end{equation}
with infinitesimal viscosity $\nu=a(D_r+D_l)$, single-particle velocity
$\tilde{v} = D_r-D_l + a (\tilde{w}_{42} - \tilde{w}_{43})$,
nonlinearity $\tilde{\theta} = 2(D_r-D_l) + a
(\tilde{w}_{42}+\tilde{w}_{24} - \tilde{w}_{43} - \tilde{w}_{34})$
and reaction constant
$k = \tilde{w}_{42}+\tilde{w}_{24} + \tilde{w}_{43} + \tilde{w}_{34}$.
This equation was studied by Murray \cite{Murr80} where
it was shown that there are shock-like travelling wave solutions.
For $a=0$ the stationary equation reduces to an ordinary first order
differential equation. With fixed boundary densities the solution can
be obtained using the approach of Ref. \cite{Popk03}.

\subsection{Asymmetric Kawasaki-Glauber Process (AKGP)}

In this case the nonzero rates are the death and branching rates
as well as one hopping rate where without loss of generality we
consider nonvanishing hopping to the left
\begin{equation}
w_{12},\quad w_{13},\quad w_{42},\quad
w_{43},\quad
w_{32}, \quad \alpha, \quad \beta \label{deathfront}
\end{equation}
In the absence of diffusion ($w_{32}=0$) this model is a variant
of zero-temperature Glauber dynamics \cite{Glau63} with a non-vanishing
magnetization current \cite{Khor01}. Including diffusion corresponds to a
nonequilibrium coupling of the zero-temperature process to an
infinite-temperature heat bath with asymmetric Kawasaki spin exchange dynamics.
In a biological
context branching corresponds to cell duplication by mitosis
which can occur only if there is space available for a second cell. In this
setting the death process describes the effect of certain types of drugs which
kill both cells in the event of mitosis \cite{Dras96}.
The two stationary densities are 0 and 1 respectively.
The diffusive shock measures are of the
same form as for the ASEP and BCRW resp. with $\rho_1=1$, $\rho_2=0$.
Given an initial
step function profile with a single domain wall ...1111100000... it is clear
that the only events that can occur are the hopping of the domain wall to
the right or left. Thus the domain wall performs a lattice random walk
just as in the previous examples with hopping rates
\bel{3-3a}
d_1 = w_{13}, \quad d_2 = w_{43}
\ee
to left ($d_1$) and right ($d_2$) respectively.

We remark that
for a special choice of the branching rates $w_{43} = w_{12} - D_l$,
$w_{42} = w_{13} + D_l$ the nonlinear contributions to the equations
of motion (\ref{5-1}) vanish identically \cite{Schu95}. Hence the
{\it exact} evolution of the density profile is given by a lattice
diffusion equation for all initial distributions. In this
case the equations of motion for the density $\rho(x,t)$ do not give any
indication of the existence
of stable shocks. Hence the
existence of a non-linearity in the dynamical equation for
the density is not necessary for having shocks in the associated
process.

\section{ASEP with open boundaries}
\label{ASEP}

Here we wish to explore some of the ramifications of the results of the
previous section on shock diffusion in the ASEP with open
boundaries.
\begin{itemize}
\item[1)]
As discussed above the stationary distribution of the shock position
describes the steady state properties of the exclusion process with
open boundaries along the manifold of boundary parameters defined by
(\ref{asepfront2})-(\ref{asepfront3}). The exact steady state properties of
the exclusion process for all values of the boundary parameters may
be calculated explicitly either by solving recursion relations \cite{Schu93}
or by using the so-called matrix product approach \cite{Derr93} which
involves the representation theory of a quadratic algebra equivalent
to a $q$-deformed harmonic oscillator algebra \cite{Sand94}.
The conditions (\ref{asepfront1})-(\ref{asepfront3}) are equivalent to
the conditions for the existence of a two-dimensional representation of the
Fock-representation of the quadratic algebra used to calculate
the stationary state properties of this model in Refs.
\cite{Derr93,EssRit1996,Mall97}. This can be seen as follows:
Let us define the function $\kappa_+$ as:
\begin{equation}
\kappa_+(x,y) = \frac{-x+y+w_{23}-w_{32}+
         \sqrt{(-x+y+w_{23}-w_{32})^2 + 4 xy}}{2x}
\end{equation}
Then the conditions (\ref{asepfront1})--(\ref{asepfront3}) can be written
in the form:
\begin{eqnarray}
\rho_1 &=& \frac{1}{1+\kappa_+(\gamma,\alpha)} \\
\rho_2 &=&
\frac{\kappa_+(\delta,\beta)}{1+\kappa_+(\delta,\beta)}  \\
w_{23} &=& 
\kappa_+(\gamma,\alpha) \kappa_+(\delta,\beta) 
w_{32}
\end{eqnarray}
We remind the reader that higher dimensional representations satisfy 
\bel{high}
(w_{23}/w_{32})^n = 
\kappa_+(\gamma,\alpha) \kappa_+(\delta,\beta),
\ee
for the derivation see \cite{Mall97}.

\item[2)]
Diffusive shock measurees have also been considered 
for the infinite system. This was done in \cite{Pigo00} for discrete time
evolution (parallel updating) and in \cite{Beli02} for continous time 
evolution. The mathematical structure behind the one-particle nature of
the bulk shock motion is the $U_q[SU(2)]$-symmetry of the generator of
the ASEP with reflecting boundaries which relates one-particle states
to shock states through the action of the ladder operator of $U_q[SU(2)]$.
Surprisingly the open boundary conditions considered here break the symmetry,
Yet the single-particle nature of the shock remains. This is reminiscent
of partially broken symmetries observed in spin chains with diagonal
boundary fields \cite{Ritt91}.

\item[3)]
 In the infinite system consecutive multiple shocks which
each satisfy (\ref{asepfront1}) evolve according to $n$-particle
dynamics, i.e., the $U_q[SU(2)]$-symmetry relates these $n$-shock measures
to states with $n$ particles \cite{Beli02}. Using the ansatz discussed
above it is easy to verify that by imposing the boundary 
conditions (\ref{high}) required
for the existence of $n$-dimensional representations of the stationary
quadratic algebra one also obtains closed equations of motion for shock
measures in the open system. To see this one adopts a slightly different
definition of the shock measure introduced in \cite{Beli02}: It is a product 
measure with density 1 at the shock positions $k_i$ and intermediate densities 
$\rho_i$ between sites $k_{i-1},k_i$. The consecutive densities each satisfy 
(\ref{asepfront1}) which by iteration leads to the
condition (\ref{high}) for the existence of $n$-dimensional representations of 
the 
stationary algebra. Hence
the $n$-dimensional representations of the stationary algebra describe
the stationary linear combination of shock measures with $n$ consecutive
shocks. Notice that this modified definition of shock measures also
allows for a representation of arbitrary shock measures (not satisfying
any special relation between consecutive densities) in terms of linear
combinations of the special shock measures.

\item[4)]
The definition of the shock position by the jump property of the shock measure
(or the presence of a particle with probability 1 at the shock position
in the alternative definition, respectively)
is not a microscopic definition of the shock position for a single realization 
of the process. As the system evolves in time one cannot trace the
shock position to tell where exactly the shock is located. In a
single realization the shock position would emerge only after spatial 
coarse graining. 
A convenient definition of the microscopic random position $X_t$ of the shock 
in a single realization of the process is the 
position of the {\it second 
class particle} \cite{Andj88,Bold89,Ferr91}.     
This particle  behaves like an ordinary particle with respect to empty sites
and like an empty site with respect to an ordinary particle.
The second class particle has a drift towards the shock \cite{Spoh91} 
and its position may thus be defined as the position of the shock.
(Cf. \cite{Derr98} for this choice.) Its diffusion coefficient has
been obtained in Ref. \cite{Ferr94}.
The density profile of the invariant shock measure as seen from the second 
class particle was calculated in \cite{Derr97}.
Its shape depends on the hopping ratio $q=\sqrt{D_r/D_l}$ and on the
limiting densities. For limiting densities satisfying (\ref{asepfront1}) 
the exact state is a pure
Bernoulli shock measure with densities $\rho_1$ to left of the second-class
particle and $\rho_2$ to its right.
This observation suggests investigating the dynamics of shock measures
with {\it second class} particles at the shock positions. 

We apply
again the strategy of following the time evolution of shock measures
by calculating the action of the Hamiltonian on the measure. In the
infinite system one finds
that indeed these measures form a closed sector analogous to $\cal U$
if the condition (\ref{asepfront1}) is satisfied for consecutive
shock densities. For the study of the open system we also need to
define the properties of the second class particle at the boundary
sites.
To this end we denote second-class particles by the symbol $B$ and
explicitly consider reservoir sites $0,L+1$ which may
either contain a $B$-particle with probability 1 or no $B$-particle, but
an $A$-particle with probability $\rho_{1,2}$ respectively. We denote
these two possible configurations of the boundary reservoir by $R_1$
and $B$ respectively and represent the left boundary processes as
follows:
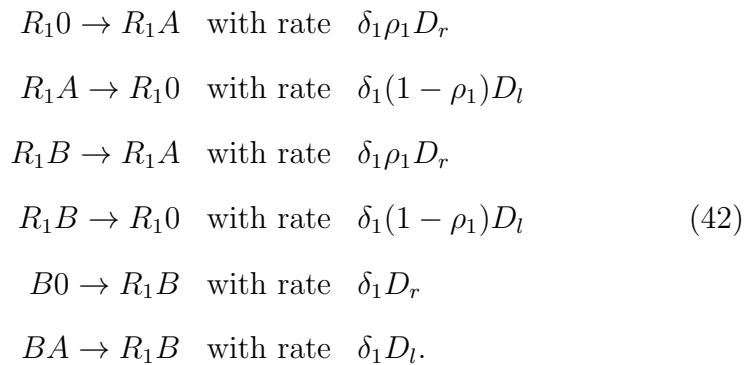
\bea
R_1 0 \to R_1 A & \mbox{with rate} & \delta_1 \rho_1 D_r \nonumber \\
R_1 A \to R_1 0 & \mbox{with rate} & \delta_1 (1-\rho_1) D_l \nonumber \\
R_1 B \to R_1 A & \mbox{with rate} & \delta_1 \rho_1 D_r \nonumber \\
R_1 B \to R_1 0 & \mbox{with rate} & \delta_1 (1-\rho_1) D_l \\
B 0 \to R_1 B & \mbox{with rate} & \delta_1 D_r \nonumber \\
B A \to R_1 B & \mbox{with rate} & \delta_1 D_l \nonumber.
\eea
At the right boundary we define 
analogously two reservoir states $R_2,B$ resp. on site $L+1$
and allow for processes with rates
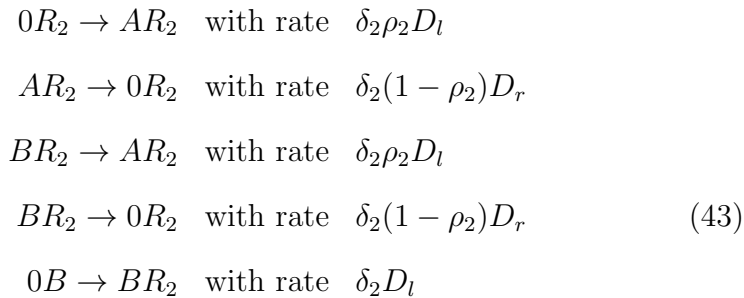
\bea
0 R_2 \to A R_2 & \mbox{with rate} & \delta_2 \rho_2 D_l \nonumber \\
A R_2 \to 0 R_2 & \mbox{with rate} & \delta_2 (1-\rho_2) D_r \nonumber \\
B R_2 \to A R_2 & \mbox{with rate} & \delta_2 \rho_2 D_l \nonumber \\
B R_2 \to 0 R_2 & \mbox{with rate} & \delta_2 (1-\rho_2) D_r  \\
0 B \to B R_2 & \mbox{with rate} & \delta_2 D_l \nonumber \\
A B \to B R_2 & \mbox{with rate} & \delta_2 D_r \nonumber .
\eea
The number of $B$-particles is conserved in this dynamics. Physically
this corresponds
to the reflection of shocks at the boundaries in the open system.
The first two transition rules for both boundaries which do not involve
$B$-particles satisfy the
injection/absorption rules (\ref{asepfront2}), (\ref{asepfront3}).

Considering the system with several $B$-particles
one may study the time evolution of $n$ consecutive shocks
with increasing densities at each point of discontinuity.
For multiple shock measures with consecutive densities satisfying
\be
\frac{D_r}{D_l} = \frac{\rho_{i+1}(1-\rho_i)}{(1-\rho_{i+1})\rho_{i}}
\ee
one finds $n$-particle dynamics.
On the hydrodynamic scale $n$ consecutive shocks obey simple deterministic
$n$-body dynamics: They move with constant speed until two shocks meet
and then coalesce into a single shock. Thus after some time only one
shock (the leftmost, which is the fastest) survives \cite{Vares}.
For the special family of boundary densities
considered above this phenomenon can be studied on the lattice scale.
\end{itemize}

\section{Conclusions}

We have studied the dynamics of a class of reaction-diffusion processes
with open boundaries on the {\it lattice
scale} and established a complete list of models
where exact travelling-shock solutions exist. For these systems we have
detailed
knowledge about the microscopic structure of the shock.
We found that there a three families of such models:
The ASEP, the BCRW, but on a more general manifold of parameters as
considered previously, and a Kawasaki-Glauber spin-flip dynamics. In all
three cases the time evolution
of the shock measure is equivalent to that of a
random walker on a lattice with $L+1$ sites
with homogeneous hopping rates in the bulk and special reflection
rates at the
boundary. The existence of such processes
imply a rather remarkable property. Shocks behave
like collective single-particle excitations already on the lattice scale --
not only after coarse-graining where all the microscopic features of the
shock are lost. The reduction of the exponentially large number of microscopic
internal degrees of freedom ($2^L$) to an only polynomially large number of
macroscopically relevant
degrees of freedom ($L+1$)
is not an uncontrolled and only phenomenologically motivated approximation,
but an exact result on all scales of observation.

As is long known from zero-temperature Glauber dynamics a hydrodynamic
description of an interacting particle system in terms of a PDE for the
particle density
may fail to give any hint on the microscopic structure of the
macroscopic solution even if the hydrodynamic equation is exact.
Our results for the AKGP indicates that this property is not a special
feature only of Glauber dynamics. It remains as an open question
under which general conditions and in which way the presence of a stable shock
in a stochastic interacting particle system is reflected in the structure of
the hydrodynamic limit. It also would be interesting to investigate travelling
shocks in systems with defects, where in the case of the ASEP exact
results are available for the steady state \cite{Schu93a,Mall96,Lee97,Jafa00}.

\newpage

\bibliographystyle{unsrt}

\end{document}